%% file: CCD4ANNS.tex
\documentclass[conference]{IEEEtran}
\IEEEoverridecommandlockouts

\usepackage{subcaption}
\usepackage{cite}
\usepackage{amsmath,amssymb,amsfonts}
\usepackage{algorithm}
\usepackage{algpseudocode}
\usepackage{graphicx}
\usepackage{textcomp}
\usepackage{listings}
\usepackage{xcolor}
\usepackage{inconsolata}        
\usepackage{upquote}            
\usepackage{xcolor}
\usepackage{hyperref}
\usepackage{multirow,array}
\usepackage{listings}
\usepackage{xcolor}
\definecolor{hlbg}{RGB}{255,241,168}   
\definecolor{dimfg}{RGB}{140,140,140}

\newcommand{\hl}[1]{\colorbox{hlbg}{#1}}
\newcommand{\dimtext}[1]{{\color{dimfg}#1}}

\def\BibTeX{{\rm B\kern-.05em{\sc i\kern-.025em b}\kern-.08em
    T\kern-.1667em\lower.7ex\hbox{E}\kern-.125emX}}
\begin{document}

\title{CCD-Level and Load-Aware Thread Orchestration \\for In-Memory Vector ANNS on Multi-Core CPUs}

\author{
	\IEEEauthorblockN{
		Yuchen Huang\IEEEauthorrefmark{1}\IEEEauthorrefmark{2}\IEEEauthorrefmark{3},
		Baiteng Ma\IEEEauthorrefmark{1}\IEEEauthorrefmark{2}\IEEEauthorrefmark{3},
		Yiping Sun\IEEEauthorrefmark{3},
		Yang Shi\IEEEauthorrefmark{3},
		Xiao Chen\IEEEauthorrefmark{3}\\
		Xiaocheng Zhong\IEEEauthorrefmark{3},
		Zhiyong Wang\IEEEauthorrefmark{3},
		Yao Hu\IEEEauthorrefmark{3},
		Chuliang Weng\IEEEauthorrefmark{2}
		\thanks{\IEEEauthorrefmark{1}These authors contributed equally to this work.}
	}
	\IEEEauthorblockA{
		\IEEEauthorrefmark{2}\textit{East China Normal University}, 
		\IEEEauthorrefmark{3}\textit{Xiaohongshu Inc (RedNote)} \\
		\IEEEauthorrefmark{2}\{ychuang, btma\}@stu.ecnu.edu.cn,
		\IEEEauthorrefmark{3}\{sunyiping, shiyang1, chenxiao4\}@xiaohongshu.com,\\
		\IEEEauthorrefmark{3}\{zhongxiaocheng, sunzhenghuai, xiahou\}@xiaohongshu.com,
		\IEEEauthorrefmark{2}clweng@dase.ecnu.edu.cn
	}
}

\maketitle

\begin{abstract}
Vector approximate nearest neighbor search (ANNS) underpins search engines, recommendation systems, and advertising services. Recent advances in ANNS indexes make CPU a cost-effective choice for serving million-scale, in-memory vector search, yet per-core throughput remains constrained by memory-access latency of vector reading and the compute intensity of distance evaluations in production deployments. With the growing scale of the business and advances in hardware, modern CCD-based multi-core CPUs have been widely deployed for high throughput in our services. However, we find that simply increasing core counts does not yield optimal performance scaling.

To improve the efficiency of more cores from the CCD-based architecture, we analyze the distributions of real-world requests in our production environments. We observe high access locality in vector search in our online services and low cache utilization, resulting from overlooking the multi-chiplet nature of CCD-based CPUs.
Hence, we propose a workload- and hardware-aware thread orchestration framework at CCD-level that (i) provides a uniform interface for both inter-query parallel HNSW search and intra-query parallel IVF search, (ii) achieves cache-friendly and workload-adaptive mapping of task dispatching, and (iii) employs CCD-aware task stealing to address load imbalance.

Applied to real production workloads from search, recommendation, and advertising services of Xiaohongshu (RedNote), our approach delivers up to 3.7$\times$ higher throughput and 30–90\% reductions in P50 and P999 latency. In detail, compared with the original framework, the cache-miss ratio decreases by 6–30\%, and the total CPU stall is reduced by 20–80\%.
\end{abstract}


\input{1-Intro.tex}

\input{2-Bak.tex}

\input{3-Moti.tex}

\input{4-Design.tex}

\input{5-Thread.tex}

\input{6-Index.tex}

\input{6-2.tex}

\input{7-Eval.tex}

\input{8-Related.tex}

\input{9-Conclusion.tex}

\section*{Acknowledgment}
We sincerely thank anonymous reviewers for their valuable feedback. This work was conducted by Yuchen Huang and Baiteng Ma during their internship at RedNote Engine Architecture Department. Yuchen Huang and Chuliang Weng are supported by the National Natural Science Foundation of China (Grant No.62272171). Chuliang Weng is the corresponding author.

\section*{AI-Generated Content Acknowledgement}
We gratefully acknowledge the assistance of AI models (e.g., Google’s Gemini model) in this work. AI tools supported our coding process by suggesting implementations and refactoring routines, which improved reliability and readability. We also used AI tools to proofread and correct grammar across the entire manuscript, helping us ensure clearer, more consistent expression without altering the scientific content. We sincerely thank their developing teams for enabling these capabilities. All the assistance of generative AI tools is under the authors’ supervision and final review.

\bibliographystyle{IEEEtran}
\bibliography{ref_title_protected}

\end{document}

%% file: 1-Intro.tex
\section{Introduction}
Multimodal unstructured data (e.g., images, documents, and videos) are usually embedded into high-dimensional vectors for efficient retrieval~\cite{word2vec,vedioRec,RD-Specific}. At RedNote~\cite{RedNote}, our search engine, recommendation system, and advertising services deploy thousands of million-scale vector tables with Approximate nearest neighbor search (ANNS). Moreover, with hundreds of millions of monthly active users (MAU)~\cite{RedNote-MAU}, we must sustain tens of millions of queries per second (QPS) under millisecond-level latency across these tables. While GPUs offer fast index construction via massive parallelism and high memory bandwidth~\cite{CAGRA,cuVS}, in-memory vector indexes (e.g., HNSW, Vamana, and IVF~\cite{hnsw,vamana,IVFFlat}) on CPUs are cost-effective choices for low-latency online serving to meet the actual performance requirements. 
\begin{figure}[t]
	\centering
	\includegraphics[width=\linewidth]{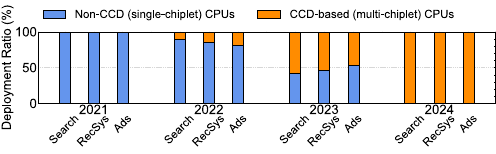}
	\vspace{-6mm}
	\caption{\textbf{\textit{The proportions of non-CCD and CCD-based multi-core CPUs deployed on services of RedNote in recent years.}}}
	\label{fig:ccdtrend}
	\vspace{-6mm}
\end{figure}

Driven by the need to push throughput under strict service-level agreements (SLA) of response latency and recall rate, our services are urgent for CPU cores. At the same time, as monolithic CPU scaling currently hits reticle-size, yield, and cost limits, vendors expand core counts of these modern CPUs through Core-Complex-Dies (CCD) architectures~\cite{AMD-9004, ccdArch}. 
Different from the scaling of cores by adding more sockets and processors as NUMA (Non-Uniform Memory Access) nodes (i.e., inter-socket scaling), CCD-based CPUs scale cores in a single socket (i.e., intra-socket scaling).

With a steady year-on-year increase in the share of these CCD-based CPU architectures in the server market~\cite{amdserver,awsandamd}, we have started to adopt these CPUs for vector search in online services recently.
Across the services of RedNote, including search engine, recommendation system, and advertising service, the ratio of CCD-based CPUs in production (shown in~\autoref{fig:ccdtrend}) has steadily increased and has now nearly displaced the previously used single-chiplet CPUs.

Unfortunately, simply adding more CPU cores with hardware upgrades does not yield optimal performance scaling in real workloads. To figure out the reasons, we analyze the real-world workload characteristics of  vector search in our industrial production environments and the hardware traits of modern CPU chiplets. We observe: (1) within sequential short time windows, queries hitting the same vector table tend to probe overlapping subsets of vectors, exhibiting strong memory locality; (2) different CCDs have the independent last level cache, therefore improper CCD–request dispatches can incur chiplet-level cache pollution and break affinity, intensifying memory contention; and (3) search overhead differs significantly across vector tables, while the na\"{\i}ve way to solve this load imbalance can incur a low ratio of cache reusing from cross-chiplet stealing.

To unleash the potential of CCD-based multi-core CPUs, we design a CCD-level and load-aware thread orchestration framework for in-memory vector ANNS. 

First, aiming at the graph-based HNSW index~\cite{hnsw} and the clustering-based inverted-file (IVF) index~\cite{IVFFlat}, which are widely used in our systems and other industrial online services~\cite{lindorm,Milvus,vexless}, we define a uniform task submission interface. It can easily replace the previous framework to enable search parallelization for both inter-query HNSW and intra-query IVF. Second, we design a task dispatcher, achieving a cache-friendly mapping between search tasks and chiplets, and a workload monitor to adapt the mapping to dynamic requests. It balances the skew memory traffic of multiple HNSW tables or clusters in IVF, minimizing the L3 cache pollution across requests. Third, we develop a CCD-aware stealing strategy to address the load imbalance of search tasks, which reduces cross-chiplet steals to minimize cache waste.

Prior studies have well optimized in-memory ANNS across multiple aspects—including search process~\cite{iQAN,gbm-ANNS,DARTH,Neos}, distance computation~\cite{vsag,Faiss-wiki}, parameter tuning~\cite{vsag,vdtuner}, and quantization compression~\cite{rabitQ,PQ}. However, to the best of our knowledge, no prior work has targeted thread orchestration optimizations for in-memory ANNS to improve search performance on modern multi-core CPU architecture and real-world industrial workloads. In this paper, we focus on the perspective of thread orchestration in vector search, achieving 1.4$\times$-3.7$\times$ throughput and reducing latency by up to 90\%. Our main contributions and key techniques can be summarized as follows:

\begin{itemize}
\item We share our deployment patterns of vector search comprehensively and describe the characteristics of  real-world workloads in our online industrial services.
\item We design a CCD-level and load-aware thread orchestration framework with a drop-in and compatible interface to integrate with inter-query HNSW and intra-query IVF.
\item We present how to balance dynamic memory traffic across search tasks in task dispatching and mapping adaptation, in order to achieve minimized cache pollution.
\item We propose to make CCD-aware stealing strategy according to the hardware topology when balancing the skew overhead of different search tasks.
\item We conduct real-world evaluations to show the efficiency, including end-to-end performance improvements and detailed gains (e.g., cache hit/miss, CPU stall).
\end{itemize}

%% file: 2-Bak.tex
\section{Background}
We introduce structures and search processes of two vector indexes widely adopted in industrial practice—graph-based HNSW and clustering-based IVF. 
We then discuss how these indexes are organized and operated in our online production services, including deployment choices with scenarios.
Finally, we present the micro architecture of modern multi-core CPUs (i.e., CCD-based multi-chiplet CPU).
\begin{figure}[t]
	\centering
	\begin{subfigure}[b]{0.45\linewidth}
		\includegraphics[width=\linewidth]{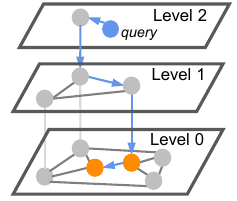}
		\caption{\textbf{\textit{Process of HNSW.}}}
		\label{fig:hnsw}
	\end{subfigure}
	\centering
	\begin{subfigure}[b]{0.45\linewidth}
		\includegraphics[width=\linewidth]{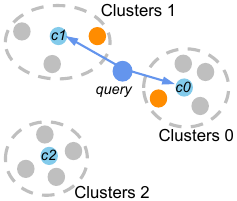}
		\caption{\textbf{\textit{Process of IVF.}}}
		\label{fig:ivf}
	\end{subfigure}
	\caption{\textbf{\textit{Structures and search processes of HNSW and IVF.}}}
	\label{fig:ablaandhyper}
	\vspace{-5mm}
\end{figure}

\subsection{Index Structure and Search}
\label{sec:index}
In this paper, we focus on the HNSW and IVF, because they are the most popular indexes for vector search and have been deployed at large scale in our production services.

\subsubsection{Graph-based HNSW}
In HNSW, each vector draws a maximum level from a geometric distribution and is inserted into all levels below. Points link up to \textit{M} neighbors to nearby ones on the same level; higher levels are sparser shortcuts, while the bottom (\textit{Level 0}) is dense and holds all vectors.

As shown in~\autoref{fig:hnsw}, starting from an entry point on the top level (i.e., \textit{Level 2}), the query performs greedy descent: at each level, it repeatedly moves to the nearest points, then drops one level. After arriving at \textit{Level 0}, a best-first search explores a candidate queue of size \textit{efSearch}, maintaining the current top-$k$. The queue stops updating when all remaining candidates cannot find nearer neighbors, and the final top-$k$ is returned.

\subsubsection{Clustering-based IVF} In IVF, vectors are partitioned into \textit{nlist} clusters via k-means; each vector is assigned to its nearest centroid and stored in the corresponding inverted list.

As shown in~\autoref{fig:ivf}, for a query, we first compute its distances to all centroids (i.e., \textit{c0, c1, c2}) and select the \textit{nprobe} closest clusters (i.e., \textit{Cluster 1 and 0}). Then, it scans the lists of nearest clusters: computing distances between query and candidates from these lists, and ranking candidates to obtain the final top-$k$.

\subsection{Deployment Choice}
Graph-based HNSW and cluster-based IVF offer complementary trade-offs between construction overhead and hardware requirements of low-latency search. 

HNSW index can minimize per-query distance evaluations, so millisecond-level latency is commonly achievable for the search process on a single core.
Its drawback is costly construction and updating. Building HNSW can take several minutes to hours at 1–100M scale and updating can take milliseconds, which makes frequent full rebuilds and inserts inefficient for workloads requiring freshness~\cite{recallbuildtimeHNSW}.

By contrast, IVF performs a coarse k-means partition followed by list assignment and can be (re)built within seconds to a few minutes and offer microsecond-level inserts, which suits high-freshness deployments~\cite{recallbuildtimeIVF}. However, query processing requires scanning many candidates across several probed lists. Achieving millisecond latency per query therefore relies on the intra-query parallelism of scanning lists and evaluating distances across multiple threads/cores (e.g., FAISS via per-list scanners and multi-threaded OpenMP~\cite{openmp}).

Hence, in practice, for services whose serving patterns favor graph indexes, we co-locate multiple HNSW indexes on a single node and execute each table’s queries on a single core to maximize system throughput. For services suited to IVF, we parallelize each query of indexes across multiple CPU cores on a node, distributing scans over the probed lists. We detail these two design choices in~\autoref{sec:prior}.
\begin{figure}[t]
	\centering
	\includegraphics[width=0.75\linewidth]{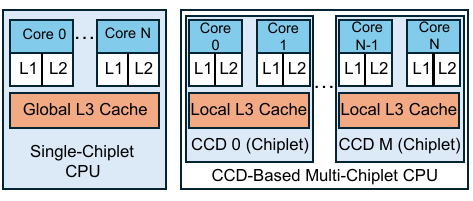}
	\caption{\textbf{\textit{Single-chiplet CPU vs. CCD-based multi-core CPU.}}}
	\label{fig:ccdarch}
	\vspace{-3mm}
\end{figure}

\subsection{CCD-based CPU Architecture}
From the perspective of hardware development, as process scaling slows, server CPUs increase core counts through multi-chiplet designs rather than monolithic dies. Recent AMD EPYC processors (e.g., Rome, Milan, Genoa, and Turin)~\cite{Milan,Genoa} and Intel’s next-generation Sapphire Rapids~\cite{intelCPU} all adopt multi-chiplet architectures. And multi-core CPUs of this type have been widely applied in our services.

As shown in ~\autoref{fig:ccdarch}, taking the AMD EPYC series as an example, multiple Core Complex Dies (CCDs) are integrated into one CCD-based multi-core CPU. Each CCD groups several cores that share a local last-level (i.e., level-3) cache (LLC/L3). The key shift from legacy single-chiplet CPUs is that the L3 cache is no longer implemented as one large, globally shared structure; instead, each chiplet (i.e., CCD) integrates its own local L3 cache, and these caches are only directly visible and low-latency to the cores within that chiplet~\cite{ccdcacheremote}. Note that CCD-based CPU scales the count of cores as a whole processor within a single socket, which is different from scaling cores by adding more sockets to equip more processors.

%% file: 3-Moti.tex
\section{Motivation}
\subsection{Prior Thread Orchestrations in Services}
\label{sec:prior}
In this subsection, we describe prior patterns of task dispatch in our online production services, including inter-query parallelism of HNSW and intra-query parallelism of IVF.

\begin{figure}[t]
	\centering
	\begin{subfigure}[b]{0.4\linewidth}
		\includegraphics[width=\linewidth]{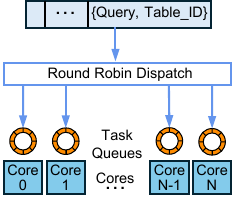}
		\caption{\textbf{\textit{Dispatch of HNSW.}}}
		\label{fig:hnsw-th}
	\end{subfigure}
	\centering
	\begin{subfigure}[b]{0.4\linewidth}
		\includegraphics[width=\linewidth]{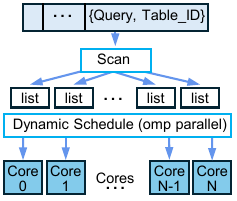}
		\caption{\textbf{\textit{Dispatch of IVF.}}}
		\label{fig:ivf-th}
	\end{subfigure}
	\caption{\textbf{\textit{Orchestration of HNSW and IVF.}}}
	\label{fig:naiveTH}
	\vspace{-3mm}
\end{figure}

\textbf{Inter-query dispatch of HNSW.} Since a single search based on HNSW typically completes within milliseconds on one core, we co-locate multiple HNSW tables on a node and drive throughput via inter-query parallelism. As shown in \autoref{fig:hnsw-th}, incoming requests—each carrying \textit{Query} (i.e., query\_vector and top-k) and \textit{Table\_ID}—are enqueued by a lightweight dispatcher to per-core ring task queues. It uses a round-robin policy that cyclically assigns the next request to the next core’s task queue, already providing O(1) scheduling overhead and avoiding centralized locks and long queues on hot cores.

\textbf{Intra-query dispatch of IVF.} For services with frequent updates and strict freshness, we adopt IVF, whose build/insert is fast but whose query is more compute-heavy than graph-based HNSW. Though the inter-query parallelism can achieve the optimal throughput for IVF indexes in theory because of less multi-thread communications than intra-query model, it can violate the latency SLAs due to the dispatch that only one core for each query. Hence, to keep millisecond latency, a single query is usually parallelized across multiple cores in our online services. As shown in \autoref{fig:ivf-th}, the query first identifies the closest posting lists; it then scans \textit{nprobe} lists in parallel (e.g., \textit{\#pragma omp for schedule(dynamic)} in FAISS). In detail, it assigns work in each of all lists that threads pull from a shared task pool as they finish a prior one, whose pull-based policy balances load approximately and utilizes all cores for intra-query parallelism.

\begin{figure}[t]
	\centering
	\begin{subfigure}[b]{0.95\linewidth}
		\includegraphics[width=\linewidth]{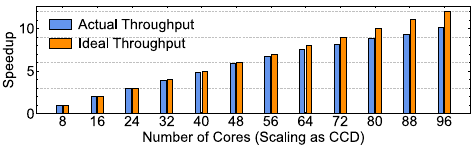}
		\caption{\textbf{\textit{Performance scaling trend of the HNSW tables.}}}
		\label{fig:hnsw-sc}
	\end{subfigure}
	\centering
	\begin{subfigure}[b]{0.95\linewidth}
		\includegraphics[width=\linewidth]{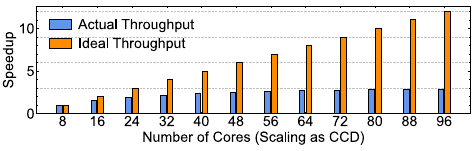}
		\caption{\textbf{\textit{Performance scaling trend of the IVF tables.}}}
		\label{fig:ivf-sc}
	\end{subfigure}
	\caption{\textbf{\textit{Scaling trends on the CCD-based multi-core CPU.}} Recall of HNSW and IVF can reach 99\% and 95\%, respectively, when requests' top-$k$ is varying between top-100-500. These are referred to workloads from our online services.}
	\label{fig:TH-SC}
\end{figure}
\subsection{Inefficiency of Scaling Cores}
\label{sec:Inefficiency of Scaling Cores}
\begin{figure*}[t]
	\centering
	\begin{subfigure}[b]{0.24\linewidth}
		\includegraphics[width=\linewidth]{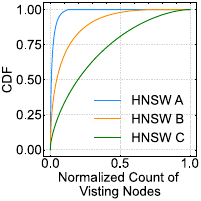}
		\caption{\textbf{\textit{Search access frequency of multiple HNSW tables.}}}
		\label{fig:hnsw-c1}
	\end{subfigure}
	\centering
	\begin{subfigure}[b]{0.24\linewidth}
		\includegraphics[width=\linewidth]{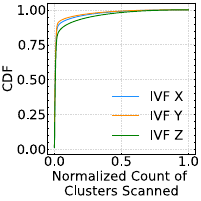}
		\caption{\textbf{\textit{Search access frequency of clusters on the IVF tables.}}}
		\label{fig:ivf-c1}
	\end{subfigure}
	\centering
	\begin{subfigure}[b]{0.24\linewidth}
		\includegraphics[width=\linewidth]{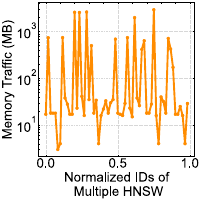}
		\caption{\textbf{\textit{Memory traffic of multiple HNSW tables.}}}
		\label{fig:hnsw-c2}
	\end{subfigure}
	\centering
	\begin{subfigure}[b]{0.24\linewidth}
		\includegraphics[width=\linewidth]{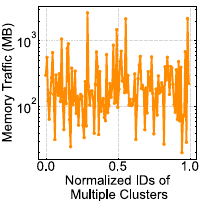}
		\caption{\textbf{\textit{Memory traffic of clusters on the IVF table.}}}
		\label{fig:ivf-c2}
	\end{subfigure}
	\centering
	\caption{\textbf{\textit{Distribution of search access frequency and memory access traffic.}}}
	\label{fig:TH-C}
	\vspace{-3mm}
\end{figure*}
Based on one of the platforms (the AMD 4th Gen EPYC 96-core processor) described in~\autoref{sec:setup}, we evaluate the scaling trends with real-world traces of our production services, including inter-query dispatch of 60 different HNSW indexes (each has 1-10 million vectors) co-located in one of our online nodes and intra-query dispatch of 15 IVF indexes (each table contains 10 thousand to 15 million vectors from one node of our frequently-updating services). The results of these two cases are shown in~\autoref{fig:TH-SC}.

We continually increase the number of CPU cores available to the vector search workload to scale overall throughput.
However, integrating these two patterns of task dispatching with the CCD-based multi-core CPU fails to deliver the anticipated improvements in throughput with scaling of cores.

~\autoref{fig:hnsw-sc} shows throughput scaling under naïve inter-query dispatch across multiple HNSW tables as the core count increases in increments of CCD-level granularity. We observe $\sim$2$\times$ to $\sim$9.9$\times$ speedup when scaling from 16 to 96 cores. However, a substantial gap remains between the actual throughput measured and ideal throughput, with the 96-core configuration achieving only $\sim$82\% of the ideal throughput (i.e., the count of CCDs $\times$ one CCD's throughput).

~\autoref{fig:ivf-sc} reports the throughput scaling of naïve intra-query dispatch in FAISS across 15 IVF tables on a single node. The trend echoes the efficient scaling of multi-table HNSW, but with poorer overall scaling. With fewer than 32 cores, there are still clear gains, whereas as the number of cores and threads increases, the incremental benefit becomes negligible. Across configurations with 2–12 CCDs, the speedup is only about 1.6$\times$ to 2.8$\times$.

These suggest that on modern CCD-based CPU architectures, adding more cores isn’t being utilized effectively in vector search. There is a substantial performance gap, which also represents an opportunity for optimization.

\subsection{Characterizing Workloads}
The ~\autoref{sec:index} presents the search workflows of both the graph-based HNSW index and the clustering-based IVF index. In HNSW, the search proceeds by following the outgoing edges of each visited node to choose the next candidate; this iterative process involves a large amount of random memory access, making this stage highly memory-intensive on latency. Similarly, in IVF, when scanning each selected cluster (list), the query must be compared against every vector in that cluster to compute distances. This requires traversing all selected lists and imposes substantial pressure on memory bandwidth.

\begin{figure}[t]
	\centering
	\includegraphics[width=0.99\linewidth]{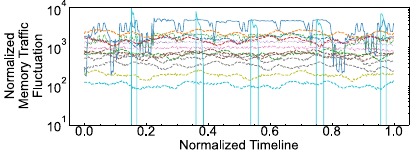}
	\caption{\textbf{\textit{Dynamic fluctuation of memory traffic along the time window.}} These are sampled requests from 15 vector tables.}
	\label{fig:TH-C3}
	\vspace{-3mm}
\end{figure}

\begin{figure}[t]
	\centering
	\begin{subfigure}[b]{0.48\linewidth}
		\includegraphics[width=\linewidth]{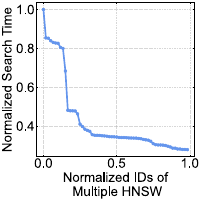}
		\caption{\textbf{\textit{Search time of multiple HNSW tables on a node.}}}
		\label{fig:hnsw-c3}
	\end{subfigure}
	\centering
	\begin{subfigure}[b]{0.48\linewidth}
		\includegraphics[width=\linewidth]{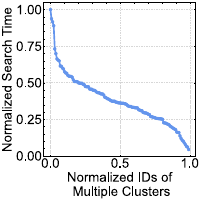}
		\caption{\textbf{\textit{Search time of multiple clusters on the IVF table.}}}
		\label{fig:ivf-c3}
	\end{subfigure}
	\centering
	\caption{\textbf{\textit{Distribution of search overhead.}}}
	\label{fig:TH-C2}
	\vspace{-3mm}
\end{figure}

Hence, to alleviate memory pressure on ANNS—both latency and bandwidth—and better harness the performance of multi-core CPUs, we next analyze and discuss the memory-access workload characteristics of in-memory ANNS in our production setting, as shown in~\autoref{fig:TH-C}, ~\autoref{fig:TH-C3} and ~\autoref{fig:TH-C2}. It is collected from online workload logs within a short time window (about one minute), and the data of vectors is the same as previous ones in~\autoref{sec:Inefficiency of Scaling Cores}.

First, the search access frequency from the single-table queries exhibits pronounced locality across both index families. Using six representative online tables (three HNSW and three IVF), we gather statistics—for each query trace within the short window—the cumulative distribution of accesses over (a) nodes visited in the HNSW graph and (b) clusters scanned in the IVF index. As shown in ~\autoref{fig:hnsw-c1} and~\autoref{fig:ivf-c1}, the CDFs rise steeply near the origin: a small fraction of nodes/clusters (lists) accounts for a large fraction of accesses during search. This heavy-tailed, Zipf-like pattern appears consistently across all six tables (HNSW A/B/C and IVF X/Y/Z), indicating that query traffic usually repeatedly touches a concentrated hotspot set rather than spreading uniformly. In short, both graph-based node visiting in HNSW and clustering-based list scans in IVF display strong search-access locality in production workloads.

Second, we quantify how memory traffic (i.e., the total volume of bytes accessed---reads, writes) is distributed across many co-located HNSW tables on a single node (as shown in  ~\autoref{fig:hnsw-c2}) and clusters within the IVF table (as shown in  ~\autoref{fig:ivf-c2}).
For HNSW, we analyze the online node that hosts $\sim$60 independent tables; for IVF, we analyze one 2M-level dataset partitioned into 8192 clusters and break memory traffic down by its clusters (lists) scanned as a representative example. The per-table/per-cluster traffic is plotted on a log scale (MB) against normalized IDs. Both panels reveal extreme skew: a small subset of HNSW tables and a small subset of IVF clusters account for disproportionately large traffic, with per-item volumes varying by one to several orders of magnitude (from tens to thousands of MB). In other words, memory accesses concentrate on “hot” tables/lists rather than spreading evenly. 
Additionally, the memory traffic of each table is dynamically changing as the requests' fluctuation. As shown in~\autoref{fig:TH-C3}, we present the fluctuations of tables' normalized memory traffic with a fixed sampling window (e.g., minute-level segment sampled here). Skew memory traffic of each HNSW table or cluster of IVF would change with the dynamic throughput, which means the “hot” and “cold” roles would frequently transmit as the sliding of time windows.

Third, we measure the per-item processing time on the same traces in a single-core run to minimize confounding factors and obtain the actual average overhead of a single search. As shown in~\autoref{fig:TH-C2}, 
these search/scan items are sorted by their mean latency. ~\autoref{fig:hnsw-c3} details the per-table search time for the 60 co-located HNSW tables, and ~\autoref{fig:ivf-c3} shows the per-cluster scan time within the single IVF table.
Both curves display a steep head and a long tail: a small subset of tables/lists incurs disproportionately high latency, while many others are comparatively light. The spread spans multiples of the median, indicating pronounced variance across items rather than uniform cost. Such skew would still directly translate into multi-core load imbalance, though we submit search tasks uniformly under naïve task dispatches (e.g., round-robin submission for multiple HNSW tables or static list chunks for scans of IVF index).

\subsection{Motivation and Challenges}

\textbf{Motivation.} \textit{Effective use of the L3 cache can provide an opportunity to alleviate memory pressure for both inter-query scaling of multi-table HNSW and intra-query scaling of single-table IVF.} On CCD-based multi-core CPUs, the L3 cache has expanded dramatically alongside core counts. The two generations of AMD CCD-architecture CPUs deployed in our production environment both provide 4 MB of L3 cache per core—up to 32/16 MB per CCD. Meanwhile, a small subset of HNSW nodes and IVF clusters accounts for most query touches. In practice, consecutive queries within short windows tend to revisit the same nodes/lists, creating a recurrent “hot set.” If this hot set is kept resident in the per-CCD local L3 cache, node-chasing in HNSW and dense scans in IVF can serve from L3 rather than DRAM, cutting latency and bandwidth demand. 

\textbf{Challenges.} In multi-chiplet CPUs, turning locality into performance improvement in industrial environments requires (i) \emph{compatibility}—a drop-in orchestration framework,
(ii) \emph{CCD-friendly task dispatching with dynamic adaptation}, and (iii) \emph{load balance with awareness of hardware topology}. 

\textit{Compatibility and ease of adoption.} It needs a minimally invasive solution. We require a unified framework in thread orchestration that exposes uniform task-submission and can keep the hardware details largely transparent while supporting inter-query HNSW and intra-query IVF without index rewrites.

\textit{CCD-friendly task dispatching with dynamic adaptation.} The L3 cache (i.e., LLC) is independent across CCDs, so naïve spreading of queries can incur cache pollution from imbalanced memory traffic. We need to map tasks to CCDs suitably, balancing dynamic hot-cold memory traffic on multiple CCDs.

\textit{Load balance with awareness of hardware topology.} Since the search overhead across HNSW tables and cluster lists in IVF is skew. The framework must react online—balancing cores for jobs and using locality-aware stealing—to meet both cache reuse and high utilization.

%% file: 4-Design.tex
\section{Design Overview}

\begin{figure}[t]
	\centering
	\includegraphics[width=0.95\linewidth]{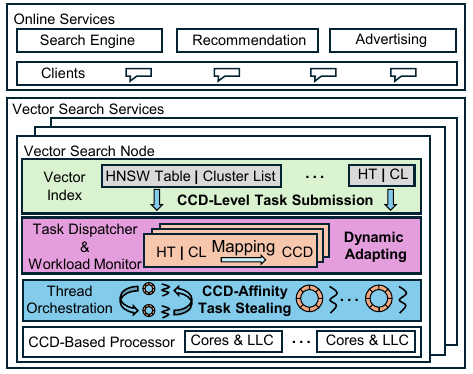}
	\caption{\textbf{\textit{Overview of our framework.}}}
	\label{fig:overview}
	\vspace{-7mm}
\end{figure}

Our goal is to turn multi-core CPUs with per-CCD cache locality into end-to-end performance improvements while keeping cores effectively utilized and source codes of indexes unchanged. As shown in ~\autoref{fig:overview}, we introduce a drop-in orchestration framework between the vector indexes (inter-query HNSW and intra-query IVF) and the CCD-based processor. The framework comprises three cooperating components: a highly compatible \emph{CCD-Level Task Submission}, a \emph{Task Dispatcher \& Workload Monitor} that performs \emph{Mapping \& Dynamic Adapting}, and a \emph{Thread Orchestration} runtime with \emph{CCD-Affinity Task Stealing}. The interface can replace a conventional thread orchestration framework used in our online services previously with uniform task submission and minimal alterations.

\textbf{CCD-level task submission.} Based on this abstraction of our framework,
codes of various indexes can emit different search task implementations to our framework via a uniform API:
\texttt{submit(search\_functor, query, Mapping\_ID)}.
Here, \emph{search\_functor} is a pluggable callable that can directly hook into
(i) an inter-query HNSW table search or
(ii) an IVF cluster-list flat scan, and returns a partial top-$k$.
The \emph{query} object encapsulates request metadata such as the desired $k$ value of the nearest neighbors' count, the raw query vector, optional filters from the original index implementations, and the source client information.
We unify identifiers by abstracting both inter-query HNSW tables and clusters of intra-query IVF tables with \emph{Mapping\_ID}. The dispatcher uses this \emph{Mapping\_ID} to package search tasks and routes tasks to CCDs while keeping the index code unchanged and the submission interface minimal.

\textbf{Task dispatcher and workload monitor for adapting mapping.}
This module provides routing of each submitted task to cores as the CCD level by maintaining a mapping from \emph{Mapping\_ID} to task queues. The dispatcher builds this mapping to meet two goals: (i) keep a HNSW table or an IVF cluster’s working set stably resident in one CCD’s LLC, and (ii) avoid concentrating multiple high-traffic keys on the same CCD, which would over-concentrate memory traffic and pollute the cache (frequently evicting the LLC footprint of HNSW tables or cluster lists). The Workload Monitor continuously measures per–table/cluster request frequency with bytes touched, and then adapts the mapping online: it remaps and spreads coincident hot keys across different CCDs adaptively, and mixes hot and cold keys evenly within a CCD. 

\textbf{CCD-aware task stealing in thread orchestration.}
Workers are pinned one-per-core and pull from local task deques. To address the load imbalance across search, we develop a CCD-affinity task stealing strategy with hierarchy.
It prefers same-CCD victims to retain LLC hits and only enables cross-CCD steals under whole-CCD idleness and sustained imbalance. The runtime supports both execution styles—independent HNSW queries and parallel IVF list scans with private per-thread top-$k$ merged on completion—thus providing compatibility across workloads while meeting locality and utilization goals.
Moreover, the processors' hardware topology is maintained inside this orchestration layer (e.g., the exact number of CCDs, the \emph{core count per CCD}, and the \emph{core-CCD mapping}).

%% file: 5-Thread.tex
\section{CCD-Level Task Submission}
\begin{figure}[t]
	\centering
	\includegraphics[width=0.95\linewidth]{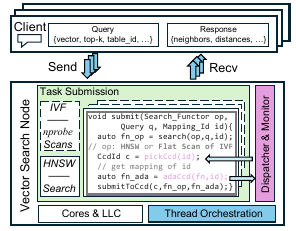}
	\caption{\textbf{\textit{Workflow of CCD-level task submission.}}}
	\label{fig:workflow}
	\vspace{-3mm}
\end{figure}

\subsection{Uniform Workflow of CCD-level Task Submission}
To minimize changes from the framework migration and ensure compatibility across different index types and parallelization modes (inter-query and intra-query),
we expose a single submission interface that both index types (HNSW and IVF) reuse:
\texttt{submit(\dots)}.
Here, \texttt{search\_functor} is a pluggable callable that binds to the
concrete search logic (either an inter-query HNSW traversal or an IVF
list scan of intra-query IVF) but remains opaque to the runtime. The \texttt{query} object
encapsulates all request metadata, including raw vector, desired top-$k$, optional
filters, and client/session information. Then, the framework can package it and
trace the task end-to-end without introducing index-specific implementation details.

We unify identifiers by treating an HNSW table and a cluster
list of an IVF table uniformly as \texttt{Mapping\_ID} (\texttt{table\_id} of HNSW and \texttt{table-cluster\_id} of IVF). CCD placement is
decoupled from the detailed implementation of different search processes: the \emph{Task Dispatcher} resolves the mapping
between
\texttt{Mapping\_ID} and the CCD where tasks are submitted to (i.e.,
\texttt{pickCcd(id)}, as shown in~\autoref{fig:workflow}), and enqueues the
task to the selected cores of CCD’s run queues. Before submission, this layer also registers the
\texttt{adaCcd(fn\_op,id)} to emit a record to the
\emph{Dispatcher \& Monitor}, enabling online workload adaptation and
periodic refinement of the mapping between CCDs and
\texttt{Mapping\_ID}. This closes the feedback loop
without changing the submission surface of our previous services or the search implementations of vector indexes.

\subsection{Inter-query and Intra-query Integration}
\textit{Integration with inter-query search of HNSW.}
Each HNSW request is packaged as a \emph{single} task that carries the
\texttt{query} with the client information. The dispatcher assigns this task to
one CCD according to its \texttt{Mapping\_ID} (i.e., its table ID). The bound
\texttt{search\_functor} executes the table search on one core of that CCD and gets
the final top-$k$ neighbors for the request; the result is returned directly to
the client. Inter-query parallelism arises from many such tasks across multiple
tables concurrently scheduled on different cores.

\textit{Integration with intra-query search of IVF.}
For each IVF table, the request first identifies the $nprobe$ nearest
clusters. The framework then \emph{decomposes} the query into multiple
per-list \emph{scan tasks}, each sharing the same \texttt{query} but
carrying a different \texttt{table-cluster\_id} as the \texttt{Mapping\_ID}. These scan
tasks are dispatched to multiple cores across cores and CCDs. Each scan task returns the local optimum top-$k$ neighbors from its cluster. Then, after all the local optimum top-$k$ are returned,
it performs a $k$-way merge, and returns the aggregated top-$k$ to the
client. This per-list scan of IVF shares the same \texttt{submit(\dots)} abstraction and delegates CCD selection, monitoring, and adaptation to the dispatcher, with the per-table HNSW search,
yielding a uniform workflow for inter-query HNSW and intra-query IVF.

%% file: 6-Index.tex
\section{Task Dispatcher and Workload Monitor}

In this section, we detail the task dispatcher and workload monitor: how to map HNSW tables and clusters of IVF tables to CCDs, and modify the dynamic adaptation that updates the mapping according to the real-time memory traffic.

\subsection{CCD Mapping of Task Dispatcher}
\label{sec:dispatcher-mapping}

\textbf{Motivation from traffic skew.}
On a single online node, when multiple HNSW tables are co-located or when all IVF tables are decomposed into clusters, the memory traffic (total bytes accessed by reads/writes during task execution) is highly skewed, as shown earlier in~\autoref{fig:hnsw-c2} and ~\autoref{fig:ivf-c2}. Some hot HNSW tables and hot clusters in IVF account for disproportionately large volumes, while other tables (or clusters) are relatively cold with fewer requests and lighter memory traffic. This skew implies that cache residency---and therefore end-to-end performance---is dominated by where these tables' and clusters' tasks are dispatched.

\textbf{Per-CCD local cache affinity.}
On CCD-based multi-core CPUs, each CCD owns a private local L3 cache shared only by the cores within that CCD. Consequently, cache affinity is created at the CCD granularity: tasks issued for the same hot table/cluster benefit when they are dispatched to the same CCD, whose L3 cache can retain the corresponding frequently-accessing memory zone.
\begin{figure}[t]
	\centering
	\includegraphics[width=0.75\linewidth]{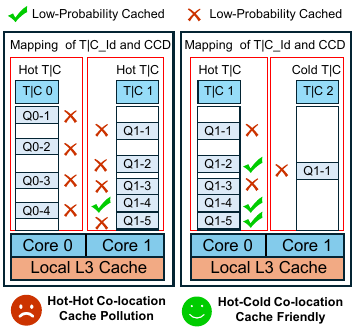}
	\caption{\textbf{\textit{Cache affinity of hot-hot and hot-cold dispatch.}} T\textbar C 0 and 1 are two hot HNSW tables (or clusters in IVF) with more queries and heavier memory traffic. T\textbar C 2 is a cold HNSW table (or cluster in IVF) with fewer queries and lighter memory traffic. Q\textit{x}-\textit{y} is the \textit{y}th query on T\textbar C \textit{x} sequentially.}
	\label{fig:hotcold}
	\vspace{-3mm}
\end{figure}

\textbf{Hot–hot vs. hot–cold and the implication for CCD mapping.}
Given the strong memory traffic skew, the mapping decision made at task submission
(i.e., \texttt{pickCcd(T\textbar C\_Id id)}) should be load-aware to be cache friendly. 
If the mapping places two \emph{hot} tables/clusters on the same CCD, then all
subsequent search tasks carrying those IDs are routed to that CCD; their bursts
interleave in a single per-CCD L3 cache, causing reciprocal evictions and cache
pollution and, in turn, lower hit rates and worse performance
(i.e., \emph{hot–hot co-location}, as shown in the left of ~\autoref{fig:hotcold}). 
Conversely, mapping a hot item together with a \emph{cold} one on a CCD
(i.e., \emph{hot–cold co-location}, as shown in the right of ~\autoref{fig:hotcold}) spreads memory traffic: the cold item contributes fewer queries in the short time window and
little L3 cache demand, allowing the hot item’s working set to remain resident and
improving in-cache retention.

\textit{Design rule for mapping.}
Hence, \texttt{pickCcd(id)} consults the current load-aware mapping in our task dispatcher
 with two priorities: (i)  maintain \emph{stickiness} so repeated submissions for the same
 T\textbar C\_id (i.e., \texttt{Mapping\_ID}) return to the same CCD to exploit memory locality; (ii) \emph{avoid} hot–hot placements on the
same CCD and \emph{prefer} hot–cold pairings to decrease cache pollution from frequent eviction.

\subsection{Dynamic Adaptation of Workload Monitor}
\label{sec:dyn-adapt-monitor}
However, neither \textsc{HNSW} nor \textsc{IVF} exposes a primitive that directly reflects the memory traffic during searches. Moreover, in an online service, this quantity fluctuates with dynamic requests and temporal locality. To address these issues, we first define an online, low-overhead estimation of per-query memory traffic. Second, we use these estimates to generate balanced hot-cold co-location to be cache friendly. Third, we develop a snapshot swap for stability, to adapt the mapping between CCDs and tables (or clusters) over time. The submission path registers \texttt{adaCcd(fn, id)} into search; upon completion, it notifies the workload monitor with measured counters (e.g., touched nodes / scanned vectors), so the monitor can update per-table (or cluster) statistics and evolve the CCD mapping.

\textbf{Approximate estimation of memory traffic.}
Let $D$ be dimensionality, $s_v$ bytes per element (e.g., 4 for FP32), and $B_v \!=\! D \cdot s_v$ the vector payload size. Let $s_{\mathrm{id}}$ be the ID width (e.g., usually 4 bytes for UINT32). For \textsc{HNSW}, a query with search width $ef_{\mathrm{search}}$ touches at most $N$ nodes (the runtime returns the exact count); each touch reads the vector and its adjacency. We approximate
\begin{equation}
	T_{\mathrm{HNSW}} \;\approx\; N \cdot \big(B_v \;+\; M \cdot s_{\mathrm{id}}\big) \;+\; \delta_{\mathrm{meta}},
\end{equation}
where $M$ is the per-node out-degree. And in practice, $\delta_{\mathrm{meta}}$ covers small write/reads for maintaining top-$k$ heap and visited list (typically $<\!1\%$) in the search process of HNSW, so it usually can be ignored.

For \textsc{IVF}, after routing to $nprobe$ cluster lists for intra-query dispatching, we attribute traffic \emph{per probed list} rather than at the whole-query level. 
For list $\mathcal{L}_i$ with $S_i \!=\! |\mathcal{L}_i|$ scanned vectors, the traffic is
\begin{equation}
	T_{\mathrm{IVF}}(\mathcal{L}_i) \;\approx\; S_i \cdot B_v \;.
\end{equation}
\begin{figure}[t]
	\centering
	\includegraphics[width=\linewidth]{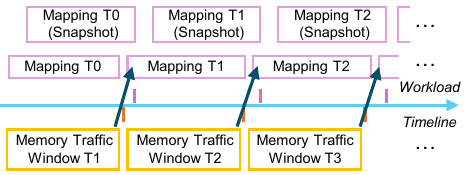}
	\caption{\textbf{\textit{Mapping adaptation with snapshot.}}}
	\label{fig:snap}
\vspace{-3mm}
\end{figure}

\textbf{Balancing CCD--item mapping with hot-cold co-location.}
Given items (i.e., HNSW tables or clusters' lists in IVF) memory-traffic estimates $\hat T_1,\dots,\hat T_n$ and $m$ CCDs, map items to CCDs so that (i) hot items are paired with cold items on the same CCD to avoid hot--hot contention, and (ii) total traffic per CCD is balanced. Let $\mu=\frac{\sum_j \hat T_j}{m}$ be the target per-CCD load.
After sorting items (tables or clusters) by traffic, we always place on the currently least-loaded CCD, and greedily pair the largest remaining item with the smallest remaining item so the placement approaches $\mu$ while encouraging hot--cold co-location. 
\begin{algorithm}[H]
	\caption{Balanced Hot--Cold Pairing for Mapping}
	\begin{algorithmic}[1]
		\Require traffic array $\hat T[1..n]$, number of CCDs $m$
		\State $\mu \gets \frac{\sum_j \hat T[j]}{m}$ \Comment{target per-CCD load}
		\State sort items by $\hat T$ (desc) into $A[1..n]$ \Comment{ordering by heat}
		\State initialize memory traffic of workloads $L[1...m]\gets 0$; \ $i\gets 1$; \ $j\gets n$ \Comment{ two-ended sweep for hot and cold memory traffic}
		\While{$i \le j$}
		\State $r^\star \gets \textsc{Get}\min_{r} L[r]$ \Comment{get ssleast-loaded CCD}
		\State $t \gets A[i]$; \ $i \gets i+1$
		\State $\text{cap} \gets \max\{0,\ \mu - L[r^\star] - \hat T[t]\}$ \Comment{residual to target}
		\If{$i \le j$ \textbf{and} $\hat T[A[j]] \le \text{cap}$}
		\State $p \gets A[j]$; \ $j \gets j-1$
		\State place $\{t,p\}$ on $r^\star$; \ $L[r^\star] \gets L[r^\star] + \hat T[t] + \hat T[p]$
		\Else
		\State place $t$ on $r^\star$; \ $L[r^\star] \gets L[r^\star] + \hat T[t]$
		\EndIf
		\EndWhile
		\State \Return Mapping $\{\text{items to CCD}\}$
	\end{algorithmic}
	\label{alg:balance hot-cold}
\end{algorithm}

Details are shown in Algorithm ~\autoref{alg:balance hot-cold}. Lines \textit{1–3} compute the target per-CCD load $\mu$ and order items by estimated traffic.
Lines \textit{4–14} perform a two-ended sweep: always place on the least-loaded CCD (line \textit{5}),
take the hottest remaining item (line \textit{6}), and steer each placement toward $\mu$ via the residual capacity (line \textit{7}).
If the coldest item fits, it is paired with the hot item (lines \textit{8–10}) to promote hot–cold co-location;
otherwise, the hot item is placed alone (lines \textit{11–13}).
Lines \textit{14–15} finish the loop and return the mapping.
This greedy pairing, combined with least-load placement, balances traffic across CCDs while reducing hot–hot co-location.

\textbf{Windowed re-mapping with snapshot swap.}
As shown in~\autoref{fig:snap}, every adaptation interval (e.g., we typically use minute-level intervals in e-commerce services, and we adjust the interval based on how frequently the traffic has changed in the past for the given business.), the workload monitor aggregates events to get memory traffic of all HNSW or IVF lists in the sliding window. As the time window goes by,
the workload monitor builds a \emph{next-map} in the background while the snapshot of \emph{current-map}, used by the task dispatcher, serves traffic. Once the new mapping is ready, it publishes an new epoched snapshot for the task dispatcher: new submissions use the new mapping immediately, while in-flight tasks finish on the current-map. When all tasks of the old epoch retire, the next-map atomically replaces the current snapshot.  
This snapshot-based handover yields stable latency during reconfiguration while continuously adapting to shifting hot-cold workloads.

%% file: 6-2.tex
\section{Thread Orchestration for CCD-Based Cores}

\subsection{Motivation and Consideration}
\label{sec:orchestr}
The heavy-tailed distribution of per-item search costs renders naive round-robin scheduling ineffective, causing significant multi-core load imbalance. We address this via a classic decentralized scheduling approach: each worker, with its threads pinned to cores, maintains a private ring-based task queue. Dynamic load balancing is then achieved through work-stealing, where idle workers procure tasks from their busy peers.
\begin{figure}[t]
	\centering
	\includegraphics[width=\linewidth]{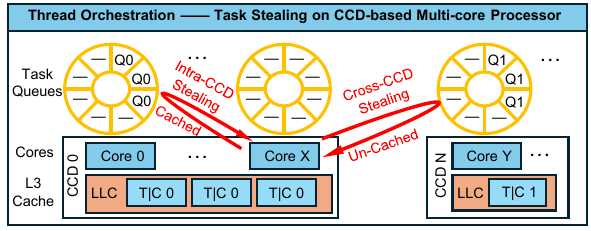}
	\caption{\textbf{\textit{Cache affinity of intra-CCD and cross-CCD task stealing.}} T\textbar C 0 and 1 in LLC mean that memory zones related to HNSW table 0 and 1 (or clusters in IVF) are cached in the L3 cache of a CCD chiplet. Q0 and Q1 represent the query on table (or cluster) 1 and 0. If Core X steals Q1 from Core Y, the memory accessing would be in the un-cached state. While if it steals Q0 from Core 0, the data can be cached.}
	\label{fig:stealing}
\end{figure}

However, on CCD-based processors, the L3 cache is private per CCD, which makes stealing locality a matter. As illustrated in ~\autoref{fig:stealing}, \emph{intra-CCD} stealing continues execution under the same LLC that already holds the hot working set (e.g., frequently revisited HNSW nodes or IVF lists), preserving cache residency. \emph{Cross-CCD} stealing migrates the task onto a different LLC; the working set is typically cold in the un-cached state there and must be re-warmed, increasing memory traffic and lowering hit rate. Hence, we need to balance skewed work while prioritizing cache affinity. 

\subsection{Topology-aware Stealing}
We prefer intra-CCD task stealing and conduct cross-CCD stealing only when necessary.
The thread orchestration module materializes the CPU topology once at initialization. For each core $i$, we record two disjoint neighbor sets: $\mathcal{S}_{\mathrm{in}}(i)$ (cores on the same CCD) and $\mathcal{S}_{\mathrm{cross}}(i)$ (cores on other CCDs). A thread is pinned to each core and follows a simple hierarchy at runtime: (1) pop locally; (2) steal within $\mathcal{S}_{\mathrm{in}}(i)$; (3) only then downgrade to attempt $\mathcal{S}_{\mathrm{cross}}(i)$. The workloop of cores is shown in Algorithm ~\autoref{alg:stealing}.

\begin{algorithm}[H]
	\caption{CCD-Topology Friendly Task Stealing Workloop}
	\label{alg:ccd-affinity}
	\begin{algorithmic}[1]
		\Require local deque $Q_i$, local CPU core $i$, intra-CCD and cross-CCD core sets of local CPU core $\mathcal{S}_{\mathrm{in}}(i)$, $\mathcal{S}_{\mathrm{cross}}(i)$
		\While{true}
		\State $t \gets \textsc{PopLocal}(Q_i)$
		\If{$t \neq null$}
		\State \textsc{Execute}$(t)$ \Comment{own-queue first to execute}
		\State \textbf{continue}
		\EndIf
		\State $t \gets \textsc{TrySteal}(\mathcal{S}_{\mathrm{in}}(i))$
		\If{$t \neq null$}
		\State \textsc{Execute}$(t)$\\ \Comment{intra-CCD steal preserves cache residency}
		\State \textbf{continue}
		\EndIf
		\State $t \gets \textsc{TrySteal}(\mathcal{S}_{\mathrm{cross}}(i))$
		\If{$t \neq null$}
		\State \textsc{Execute}$(t)$ \Comment{cross-CCD as the last choice}
		\State \textbf{continue}
		\EndIf
	\Comment{Ready to receive the next task}
		\EndWhile
	\end{algorithmic}
	\label{alg:stealing}
\end{algorithm}

%% file: 7-Eval.tex
\section{Evaluation}
In this section, we first describe the experiment setup, including environmental configurations, workloads from real-world production, and baselines we compared. Then, we present the detailed evaluations to answer the following questions:
\begin{itemize}
	\item How does the current framework we proposed affect end-to-end saturated performance on CCD-based modern CPUs, with respect to throughput and latency, compared to previous versions?
	\item To what extent are the end-to-end benefits attributable to improvements in hitting rates of L3 cache, reductions in CPU stall time, and enhanced core utilization?
	\item  How does the proposed framework perform under real-world industrial online workloads compared with the prior thread orchestration mode?

\end{itemize}

\subsection{Experiment Setup}
\label{sec:setup}
\begin{table}[htbp]
	\centering
	\setlength{\tabcolsep}{8pt}
	\renewcommand{\arraystretch}{1.2}
	\resizebox{0.49\linewidth}{!}{
		\centering
		\begin{tabular}{|c|c|c|}
			\hline
			\multicolumn{3}{|c|}{\textbf{AMD 4th Gen EPYC 96-core CPU}}\\
			\hline
			\multirow{3}{*}{\textbf{CPU}} & Core   & 96 \\ \cline{2-3}
			& CCD    & 12 \\ \cline{2-3}
			& Memory & 576GB DDR5\\ \hline
			\multirow{2}{*}{\textbf{CCD}} & Core   & 8 \\ \cline{2-3}
			& Total L3 & 32MB \\ \hline
			\multirow{4}{*}{\textbf{Core}}& L1 Cache & 32KB \\ \cline{2-3}
			& L2 Cache & 1MB \\ \cline{2-3}
			& L3 Cache & 4MB \\ \cline{2-3}
			& Frequency & 3.5GHz \\ \hline
	\end{tabular}}
	\resizebox{0.49\linewidth}{!}{
		\centering
		\begin{tabular}{|c|c|c|}
			\hline
			\multicolumn{3}{|c|}{\textbf{AMD 2nd Gen EPYC 48-core CPU}}\\
			\hline
			\multirow{3}{*}{\textbf{CPU}} & Core   & 48 \\ \cline{2-3}
			& CCD    & 12 \\ \cline{2-3}
			& Memory & 512GB DDR4 \\ \hline
			\multirow{2}{*}{\textbf{CCD}} & Core   & 4 \\ \cline{2-3}
			& Total L3 & 16MB \\ \hline
			\multirow{4}{*}{\textbf{Core}}& L1 Cache & 32KB \\ \cline{2-3}
			& L2 Cache & 0.5MB \\ \cline{2-3}
			& L3 Cache & 4MB \\ \cline{2-3}
			& Frequency & 2.6GHz \\ \hline
	\end{tabular}}
	\caption{\textbf{\textit{Hardware configuration of our platforms.}}}
	\label{tab:cpu}
\end{table}
\textbf{Hardware and software configuration.}
We use two configurations for evaluation, each equipped with an AMD multi-core CPU based on the CCD architecture (from two different generations, including AMD 4th Gen Genoa 96-core 9654 processor and AMD 2nd Gen Rome 48-core 7K62 processor). The detailed information is shown in~\autoref{tab:cpu}. The operating systems are CentOS 8 with a Linux kernel 5.10. All implementations are written in C++ and are compiled with GCC 11.4 using the -O3 optimization flag. And our HNSW and IVF indexes are implemented with reference to hnswlib~\cite{hnsw} and FAISS~\cite{IVFFlat}. The distance evaluation of both HNSW and IVF is accelerated by AVX instructions.

\textbf{Workloads.} All datasets and queries are collected from our production recommendation and advertising services. For inter-query HNSW deployment, we adopt 60 HNSW tables from a serving node, and each vector table contains 1M to 10M rows. For intra-query IVF deployment, we adopt 15 tables from a serving node, and each vector table contains 10K to 15M rows.  The vector dimensionality of both the datasets and the queries ranges from 64 to 256. And the k values of top-k are 100 to 500, and the recall rates reach above 90\% during search, which are commonly used in our services. Similarity is measured using the L2 distance. For HNSW, we set \textit{M} = 32 (i.e., number of neighbors per node) and \textit{efConstruction} = 500 (i.e., size of the candidate list during construction). For IVF, we set \textit{nlist} (i.e., number of clusters) = 128 to 8192 according to the total rows of the table, for ensuring the building time within a minute-level to meet the freshness requirements in our production services. Each \textit{efSearch} of HNSW tables is set to keep requests' recall rates reaching 99\%. And each \textit{nprobe} of IVF tables is set to keep requests' recall rates reaching 95\%. The set of queries is collected from logs of services, which contains about 1 million requests of HNSW tables and about 1.1 million requests of IVF tables. 

\textbf{Baselines.} 
We refer to the na\"{\i}ve round-robin orchestration framework as \textbf{V0 (RR)}. 
We denote as \textbf{V1 (bthread)} the thread scheduling framework that does not account for the CCD topology and augments scheduling with work stealing (following the bthread~\cite{bthread} stealing implementation), which has been widely deployed in our production workloads. 
We refer to the optimal, CCD-level and load-aware thread orchestration framework presented in this paper as \textbf{V2 (CCD)}.
For \textsc{HNSW}, we compare the performance differences among V0, V1, and V2.
For \textsc{IVF}, we use the intra-query parallel implementation of FAISS (i.e., the default OpenMP-based implementation) as comparison \textbf{V0 (FAISS)}, which submits all scans of lists within each query on all CPU cores for intra-query parallelism.

\subsection{Overall Performance}
In this subsection, we use real production workloads of both HNSW and IVF to conduct offline stress testing and evaluate performance under saturated load. We report aggregate throughput at saturation and tabulate the median (P50) latency and the tail (P999) latency, which together characterize the service’s upper bound of admissible request load and quality of service (QoS).

\begin{figure}[t]
	\centering
	\begin{subfigure}[b]{\linewidth}
		\includegraphics[width=\linewidth]{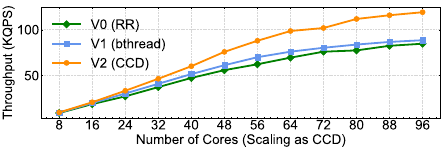}
		\caption{\textbf{\textit{Saturated throughput of HNSW on the 4th Gen 96-core CPU.}}}
		\label{fig:qpshnsw4}
	\end{subfigure}
	\centering
	\begin{subfigure}[b]{\linewidth}
		\includegraphics[width=\linewidth]{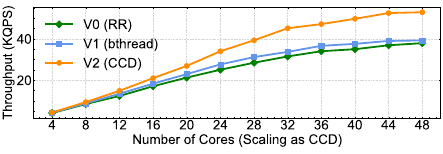}
		\caption{\textbf{\textit{Saturated throughput of HNSW on the 2nd Gen 48-core CPU.}}}
		\label{fig:qpshnsw2}
	\end{subfigure}
	\centering
	\caption{\textbf{\textit{Saturated throughput of HNSW as scaling CCDs.}}}
	\label{fig:QPS_v2}
\end{figure}
\begin{figure}[t]
	\centering
	\begin{subfigure}[b]{\linewidth}
		\includegraphics[width=\linewidth]{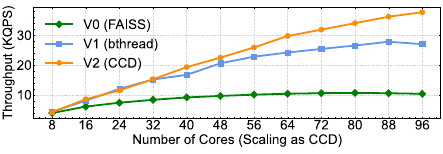}
		\caption{\textbf{\textit{Saturated throughput of IVF on the 4th Gen 96-core CPU.}}}
		\label{fig:qpsivf4}
	\end{subfigure}
	\centering
	\begin{subfigure}[b]{\linewidth}
		\includegraphics[width=\linewidth]{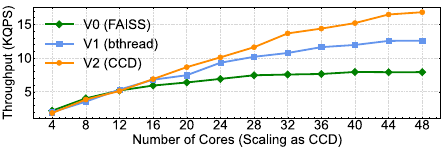}
		\caption{\textbf{\textit{Saturated throughput of IVF on the 2nd Gen 48-core CPU.}}}
		\label{fig:qpsivf2}
	\end{subfigure}
	\centering
	\caption{\textbf{\textit{Saturated throughput of IVF as scaling CCDs.}}}
	\label{fig:QPS_v1}
\end{figure}

First, we report saturated throughput on two generations of CCD-based multi-core CPUs, including the 4th Gen 96-core processor and the 2nd Gen 48-core processor. The core count is scaled one CCD at a time (i.e., at each step, we enable all cores within a CCD). As shown in~\autoref{fig:QPS_v2} and ~\autoref{fig:QPS_v1}, the CCD-aware design we proposed (V2) consistently delivers the highest throughput and scales better than the bthread-based V1 and the baseline V0. For HNSW (Fig. 14), throughput rises nearly linearly as CCDs are added, reaching over 100 KQPS at 96 cores and $\sim$50 KQPS at 48 cores, with V2 maintaining a clear margin over V1/V0 across the range. For IVF (\autoref{fig:QPS_v1}), V2 likewise shows superior scaling, attaining ~35 KQPS at 96 cores versus ~25 KQPS for V1 and ~10 KQPS for V0. These results indicate that our CCD-level and load-aware orchestration framework improves the maximum throughput and the scalability of performance on the CCD-based architecture.

\begin{figure}[t]
	\centering
	\begin{subfigure}[b]{\linewidth}
		\includegraphics[width=\linewidth]{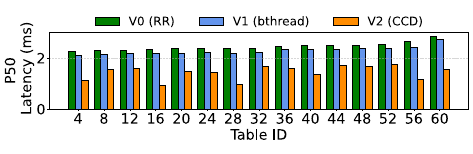}
		\caption{\textbf{\textit{Comparison of P50 search latency from HNSW tables.}}}
		\label{fig:p50ivf}
	\end{subfigure}
	\centering
	\begin{subfigure}[b]{\linewidth}
		\includegraphics[width=\linewidth]{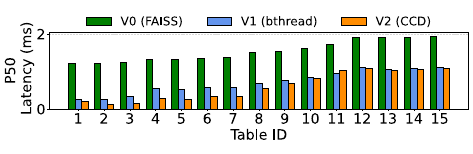}
		\caption{\textbf{\textit{Comparison of P50 search latency from IVF tables.}}}
		\label{fig:p50hnsw}
	\end{subfigure}
	\centering
	\caption{\textbf{\textit{Comparisons of P50 search latency.}}}
	\label{fig:P50}
			\vspace{-3mm}
\end{figure}

\begin{figure}[t]
	\centering
	\begin{subfigure}[b]{\linewidth}
		\includegraphics[width=\linewidth]{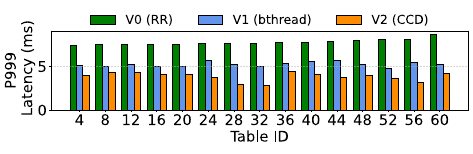}
		\caption{\textbf{\textit{Comparison of P999 search latency from HNSW tables.}}}
		\label{fig:p999ivf}
	\end{subfigure}
	\centering
	\begin{subfigure}[b]{\linewidth}
		\includegraphics[width=\linewidth]{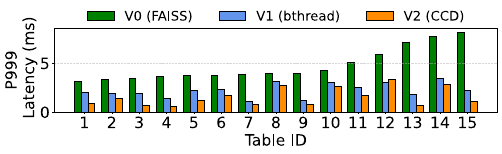}
		\caption{\textbf{\textit{Comparison of P999 search latency from IVF tables.}}}
		\label{fig:p999hnsw}
	\end{subfigure}
	\centering
	\caption{\textbf{\textit{Comparisons of P999 search latency.}}}
	\label{fig:P999}
			\vspace{-3mm}
\end{figure}

Second, we report the median P50 search latency and the tail P999 search latency, which are evaluated with 96 cores of the 4th Gen EPYC processor.
Results of median latency are shown in~\autoref{fig:P50}. And results of tail latency are shown in~\autoref{fig:P999}. For the HNSW case, we display 15 out of the 60 tables (ranked by performance of V0, selecting one table out of every four). Due to limited space here, although we only show some of HNSW tables, the overall trend across all 60 tables is the same. For the IVF case, we display all 15 tables.

\autoref{fig:P50} and \autoref{fig:P999} show that the CCD-level thread orchestration framework, V2 (CCD), achieves the lowest latency across both median (P50) and tail (P999) latency for inter-query HNSW search and intra-query IVF search in our services' workloads. Relative to the baseline V0 (FAISS), V1 (bthread) introduces proactive task stealing, which balances work more evenly across workers and reduces queueing hot spots. This improves utilization and lowers both median latency and tail latency, compared with V0 (i.e., round-robin dispatching in HNSW and OpenMP in FAISS-IVF). In addition to naive task stealing, our CCD-level and load-aware V2 framework further reduces latency by improving cache efficiency (e.g., cache-friendly request dispatching and reduced cross-chiplet task stealing), so that more query processing happens in the CPU caches rather than in slower memory. As a result, V2 we proposed in this paper consistently outperforms V1 and V0 on both HNSW and IVF setups in terms of P50 and P999 search latency.

\subsection{Detailed Study}
\begin{figure}[t]
	\centering
	\begin{subfigure}[b]{\linewidth}
		\includegraphics[width=\linewidth]{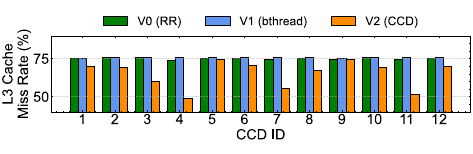}
				\vspace{-7mm}
		\caption{\textbf{\textit{Cache miss rates of HNSW tables.}}}
		\label{fig:cacheHNSW}
	\end{subfigure}
	\centering
	\begin{subfigure}[b]{\linewidth}
		\includegraphics[width=\linewidth]{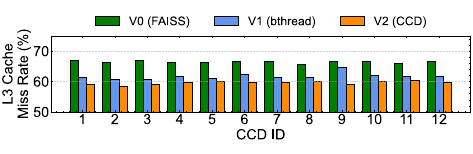}
				\vspace{-7mm}
		\caption{\textbf{\textit{Cache miss rates of IVF tables.}}}
		\label{fig:cacheIVF}
	\end{subfigure}
	\centering
	\caption{\textbf{\textit{Comparisons of L3 cache miss rates.}}}
	\label{fig:cache}
			\vspace{-3mm}
\end{figure}

\begin{figure}[t]
	\centering
	\begin{subfigure}[b]{0.56\linewidth}
	\includegraphics[width=\linewidth]{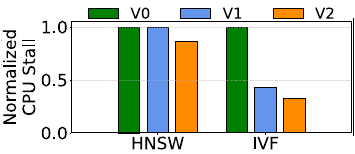}
	\caption{\textbf{\textit{Decrease of CPU stall during search.}}}
	\label{fig:stall}
	\end{subfigure}
	\centering
		\begin{subfigure}[b]{0.42\linewidth}
	\includegraphics[width=\linewidth]{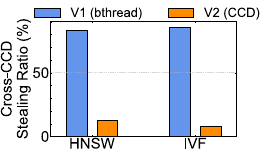}
	\caption{\textbf{\textit{Decrease of cross-CCD task stealing during search.}}}
	\label{fig:steal}
	\end{subfigure}
		\centering
	\caption{\textbf{\textit{Comparisons of CPU stall and cross-CCD stealing of both HNSW and IVF.}}}
				\label{fig:stallandsteal}
	\vspace{-5mm}
\end{figure}

In this subsection, we present the detailed report from three aspects, including overall hit/miss rates of L3 cache (~\autoref{fig:cache}), CPU stall, and the behavior of task stealing (~\autoref{fig:stallandsteal}).

First, we study the L3 cache behavior as shown in~\autoref{fig:cache}. Overall, on both HNSW (\autoref{fig:cacheHNSW}) and IVF  (\autoref{fig:cacheIVF}), our \textsc{V2} framework markedly reduces the fraction of L3–cache misses. The results are collected with \textit{AMD uProf} on the 4th Gen 96-core, 12-CCD processor while running end-to-end search workloads. For HNSW, \textsc{V2} delivers consistently lower miss rates than both \textsc{V1} (bthread) and the \textsc{V0} baseline (round-robin) across all 12 CCDs. V1 shows similar cache miss rates without optimizations addressing the CCD-based architecture. For IVF, \textsc{V1} already slightly improves upon the OpenMP-based FAISS baseline, while \textsc{V2}, benefiting from hot–cold colocation in the mapping policy and a CCD-preferential work-stealing strategy, achieves the lowest L3 cache miss rate.

Second, we report the CPU stall. As shown in \autoref{fig:stall}, we also record CPU stall under HNSW- and IVF-based search loads. Here, \emph{CPU stall} denotes cycles in which the core cannot retire instructions—typically because it is waiting on long-latency data and instruction fetches, (e.g., cache misses spilling to memory), or CPU contention (e.g., thread switches), thereby reducing effective utilization. The lower stall implies better CPU use, whereas higher stall leads to latency spikes and throughput losses. On HNSW, \textsc{V0} and \textsc{V1} show comparable stall, while \textsc{V2} achieves a noticeably lower stall thanks to fewer long-latency memory accesses arising from improved cache affinity. On IVF, \textsc{V1} benefits from task-stealing–based scheduling and already reduces stall relative to the OpenMP baseline, and \textsc{V2} further increases cache hit likelihood on top of that, yielding the lowest stall overall.

Third, we record the behavior of task stealing, when each CPU core's own task queue is empty. Results are shown in~\autoref{fig:steal}. While work-stealing is crucial for load balancing, unrestricted stealing can lead to significant performance degradation due to local L3 cache misses. Specifically, when a core in one CCD steals a task from another CCD, the subsequent execution of that task is likely to access data previously residing in the remote CCD's L3 cache and waste the cache space.
We propose to prioritize stealing tasks from cores within the same CCD when a core becomes idle. This prioritization is achieved by leveraging CPU hardware topology information to define the stealing targets' vicinity. For HNSW, the cross-CCD stealing ratio is reduced from $\sim75\%$ in V1 to below $10\%$ in V2.
For IVF, the ratio is similarly suppressed from above $80\%$ in V1 without CCD-aware to a mere $\sim5\%$ in current V2.

\subsection{Serving Performance}
\begin{figure}[t]
	\centering
	\begin{subfigure}[b]{\linewidth}
		\includegraphics[width=\linewidth]{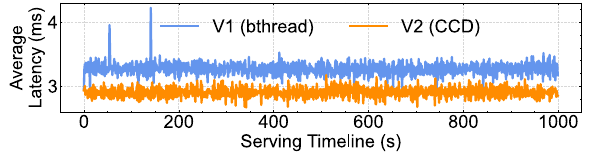}
		\caption{\textbf{\textit{Average response time of requests on HNSW tables.}}}
		\label{fig:TLHNSW50}
	\end{subfigure}
	\vspace{3mm}
	\centering
	\begin{subfigure}[b]{\linewidth}
		\includegraphics[width=\linewidth]{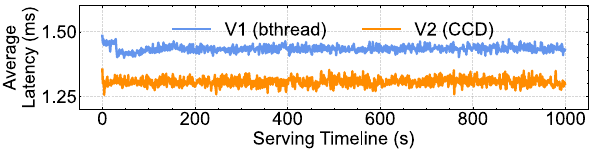}
		\caption{\textbf{\textit{Average response time of requests on IVF tables.}}}
		\label{fig:TLIVF50}
		\vspace{-3mm}
	\end{subfigure}
	\centering
	\caption{\textbf{\textit{Comparison of average response time as a timeline.}}}
	\label{fig:TL50}
			\vspace{-3mm}
\end{figure}

In the production-style pressure-limited setting, we compare two serving frameworks (V1 and V2) over a continuous 1000s run. For every second, we aggregate all requests that arrived in that second and record the average response latency. These per-second averages form the time series in ~\autoref{fig:TL50}, with HNSW and IVF backends shown in ~\autoref{fig:TLHNSW50} and ~\autoref{fig:TLIVF50}. Across both index types, V2 (CCD) delivers consistently lower average latency and tighter dispersion than V1 (bthread). On HNSW, V1 exhibits intermittent spikes, whereas V2 remains flat and stable, suggesting better Qos. On IVF, both traces are steady, yet V2 persistently tracks below V1. Overall, under realistic admission control, CCD-level and load-aware V2 provides more stable service and lower average response time. In addition, we set the time window as 10 s for the dynamic remapping of tasks here. It can be seen that the remapping strategy is a low-overhead adjustment and harmless for the services. 

%% file: 8-Related.tex
\section{Related Works}
\textbf{Research development in vector search.}
Early systems relied on locality-sensitive hashing (LSH) and tree-based indexes (e.g., KD-tree)~\cite{lsh,kdtree}, whose search efficiency degrades in high dimensions. Subsequently, partitioned (clustering) indexes (e.g., IVF and balanced $k$-means trees, BKT)~\cite{IVFFlat,sptag} and graph-based indexes (e.g., NSG, HNSW, and Vamana) deliver superior recall–latency trade-offs and become mainstreams for deployments~\cite{nsg,hnsw,vamana}. Meanwhile, quantization~\cite{PQ} advances from PQ and SQ toward more effective schemes such as the recent RaBitQ~\cite{rabitQ}, which shrink index footprints and enable shorter-code distance estimation. We are introducing these advanced quantization methods into services.
We argue that, with memory touches cut from quantization and the scaling of L3 cache from CCD increased, the effectiveness of our framework and the future finer-grained task–thread orchestration would amplify per-CCD cache benefits.

\textbf{Industrial development in vector search.}
Recent industrial practice converges on DB-centric, cloud-native designs, purpose-built or vector-augmented DBMS (e.g., Milvus, Pinecone, SingleStore\mbox{-}V) \cite{Milvus,Guo2022Manu,pinecone,Chen2024SingleStoreV}. Many existing challenges and open problems are well discussed and summarized in ~\cite{10.1145/3626246.3654691,VLDB2024Panel}.
Then, production systems in large vendors, such as BlendHouse, GaussDB\mbox{-}Vector, and Azure Cosmos~DB, present how to integrate vector search into their well-developed databases~\cite{Niu2025BlendHouse,Sun2025GaussDBVector,Upreti2025CosmosDB}. VSAG describes their framework's optimizations about graph-based ANNS comprehensively~\cite{vsag}. 
After SPANN, DiskANN, and Starling are proposed~\cite{DiskANN,SPANN,starling}, ~\cite{Shim2025TurboSSD} pgvector and Cosmos~DB show the integration of persistent vector search and relational and NoSQL databases~\cite{Sun2025GaussDBVector}. TigerVector integrates vector search and graph query within the native graph database, TigerGraph~\cite{TigerVector2025}. From another special aspect that is different from these remarkable works, we characterize the in-memory vector search of industrial services at RedNote, and propose our thread orchestration to optimize it according to hardware features of CCD-based processors.

\textbf{Research about optimization on CCD-based multi-core CPUs.} 
Modern multi-core  CPUs partition cores and L3 cache across dies (e.g., CCD on AMD EPYC), introducing non-sharing L3 cache even within a socket. 
AMD’s industry retrospectives detail the motivations and design choices behind chipletization (Infinity Fabric topology, CCD organization) and how advanced 2D/3D modular packaging co-evolves with software stacks~\cite{Milan,smith2024ehp}. 
Recent system studies reveal that hardware-oblivious software design can severely underutilize increased cores and cache capacity. And \cite{ccdolap} demonstrates that chiplet-aware placement of multiple workers (instantiations) can speed up OLAP workload up to 7~$\times$ by deploying query engines with CCD-local granularity. In the inference of recommendation, CCD-based CPUs are widely used as well. ~\cite{ccdArch} and ~\cite{ccdasplos} study how to break  bottlenecks in the deep learning model's embedding stage on CPUs by improving data reuse in caches of CCD-based architectures. These works motivate our co-design to study the real-world workloads of vector search, and develop this thread framework for services of vector search with better performance than the previous thread orchestration framework on CCD-based multi-core CPUs.

%% file: 9-Conclusion.tex
\section{Conclusion}
Driven by the requirements of large-scale online ANN services with tight latency SLA at RedNote, we analyzed why merely adding cores on modern CCD-based CPUs fails to translate hardware upgrading into throughput: short-window skew, chiplet-isolated L3 cache, and per-table/cluster imbalance jointly degrade cache residency and utilization. We presented a CCD-level and load-aware orchestration framework that unifies inter-query (HNSW) and intra-query (IVF) parallelism behind a drop-in submission interface in our services. It couples cache-friendly task-to-CCD mapping with online adaptation, and employs topology work stealing to balance load without sacrificing locality. Evaluated by real-world workloads and deployed in production, our approach delivers higher saturation throughput, reduces P50 and P999 latency to achieve better QoS stability. These results demonstrate that our co-design thread orchestration with CPU topology of the CCD architecture is an effective and practical lever for accelerating in-memory vector search in industrial services.